\documentclass[twoside]{article}
\usepackage{fleqn,espcrc2}
\usepackage{epsf}

\title{
\vspace*{-1.2cm}
\begin{flushright}
{\normalsize UTHEP-481, UT-CCP-P-145, KEK-CP-147}\\
\end{flushright}
Lattice QCD on Earth Simulator\thanks{
presented at {\it Lat'03} by T.~Yoshi\'e}}

\author{
S.~Aoki\rlap,%
\address{Institute of Physics, University of Tsukuba,
	Tsukuba, Ibaraki 305-8571, Japan}
K.-I.~Ishikawa\rlap,$^{\rm a,}$%
\address{Center for Computational Physics, University of Tsukuba,
	Tsukuba, Ibaraki 305-8577, Japan}
Y.~Iwasaki\rlap,$^{\rm a}$
K.~Kanaya\rlap,$^{\rm a}$
T.~Kaneko\rlap,%
\address{High Energy Accelerator Research Organization (KEK),
	Tsukuba, Ibaraki 305-0801, Japan}
Y.~Kuramashi\rlap,$^{\rm c}$
N.~Tsutsui\rlap,$^{\rm c}$
A.~Ukawa$^{\rm a,b}$ and
T.~Yoshi\'e$^{\rm a,b}$
}

\begin{document}

\begin{abstract}
We report on coding and performance of our polynomial hybrid Monte Carlo 
program on the Earth Simulator.  
At present the entire program achieves 25--40\% efficiency. 
An analysis of overheads shows that a tuning of 
inter-node communications is required for further improvement.

\end{abstract}

\maketitle

\section{Introduction}
The joint CP-PACS/JLQCD Collaborations have been 
carrying out~\cite{ref:Lat03-Nf3} three-flavor full QCD simulations 
with the Iwasaki RG gauge action and the non-perturbatively $O(a)$
improved Wilson quark action using 
the polynomial Hybrid Monte Carlo (PHMC) method~\cite{ref:PHMC}
on a variety of computers.
One of the computers is the Earth Simulator(ES) at the ES 
Center~\cite{ref:ES}.  
It was made available 
under the project ``Study of the Standard Model of Elementary 
Particles on the Lattice with the Earth Simulator'' which was approved  
as one of the ``Epoch Making Simulation Projects'' of the ES Center . 
Here we report on the performance of our PHMC program on ES. 

\section{Coding on ES}\label{sec:coding}
The ES consists of 640 processing nodes (PN)
connected by a one-dimensional crossbar switch with 12.3 GB/s bi-directional 
bandwidth. Each PN is an SMP with 8 vector-type arithmetic 
processors (AP), each with a peak speed of 8GFlops.
Among several programming models, we employ micro-tasking 
by hand parallelization for 8 AP's of a single PN and MPI 
for communications between PN's.

Our PHMC code was originally developed at KEK for Hitachi SR8000
and sustains as much as 40\% of peak speed as a whole.  
On the ES, however, the performance turned out to be of order 10\%. 
To explore an effective coding style, we take the basic 
MULT subroutine for Wilson-Dirac matrix-vector multiplication, 
and measure the performance for seven types of coding 
on a single AP. Results range 
from 10 to 73\% for a $6\times 6\times 48\times 48$ lattice.
 
The highest performance is achieved by a code written 
originally for a vector-parallel machine, Fujitsu VPP500. 
In this code, sites are one-dimensionalized on the $z-t$ plane and 
divided by four to realize a large vector length without list 
vectors. Other codes with smaller vector lengths (e.g. that for
Hitachi SR8000 in which only $t$ direction is vectorized)
show low performance of at most 30\%.
Hence we rewrite the entire code with the coding style
used for VPP500. 

We subdivide the whole lattice on the $x-y$ 
plane and assign each region to a PN and parallelize with MPI. 
In one PN, loop indices in the $x$ and $y$ directions
and an even/odd flag for four vector loops mentioned above are 
one-dimensionalized and divided by eight.  This enables the compiler
to do an automatic parallelization (micro-tasking) 
of all appropriate do-loops. 

\section{Performance of MULT}\label{sec:MULT}

Since performance depends on the lattice size and 
the number of PN, we examine three cases, 
a) $20^3\times N_t$ on $5\times 2$ PN 
($4\times 10\times 20\times N_t$ per PN),
b) $24^3\times N_t$ on $3\times 4$ PN
($8\times 6\times 24\times N_t$ per PN), and
c) $32^3\times N_t$ on $4\times 4$ PN
($8\times 8\times 32\times N_t$ per PN).

\subsection{Vector Processing on single AP}
\begin{figure}[t]
\centerline{\epsfxsize=6.7cm \epsfbox{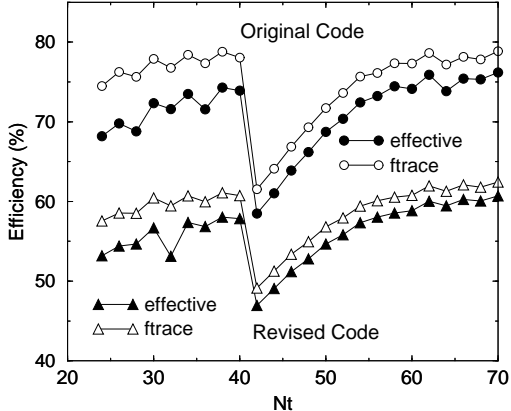}}
\vspace{-12mm}
\caption{MULT vector performance on 1AP for 
an $8\times 6\times 24\times Nt$ lattice.}
\label{fig:MULT-1AP-24}
\vspace{-6mm}
\end{figure}

The vector processing performance on a single AP is an important fundamental.
Our program includes redundant arithmetic calculations 
at edges in the $z$ direction to realize a long vector length.
Therefore we distinguish the performance reported 
by the system analyzer ```ftrace'' and that calculated
theoretically for an effective part excluding the redundant operations.
For the latter, total flops equals 1296flops$\times$\#sites.
The redundant part costs $2-4$\% of peak performance.

Figure~\ref{fig:MULT-1AP-24} shows the single AP performance of MULT
for case b) plotted versus $N_t$. We test two codes. 
In the original one, contributions from 8 directions 
are calculated in one large do-loop and are summed up later.  
In the revised code, which intends to overlay arithmetic operations
and communications in future, the large loop is divided into 
two loops for $z,t$ and $x,y$ directions. The array structure
is also different. The revised code runs about 15\% slower.  
We suppose that this is partly caused by a slow
startup of do-loops. 

In general, the vector performance of the revised MULT code 
reaches 55-65\% for all three cases. However, it drops by about 
10\% when the vector length 
just exceeds a multiple of the size of vector registers, 256.  

%From these experiences, we conclude, as usual for vector 
%processors, that do-loop structure including vector length
%should be tuned property to achieve high performance on the ES.

\subsection{Micro-Tasking Parallelization}
\begin{figure}[t]
\centerline{\epsfxsize=6.7cm \epsfbox{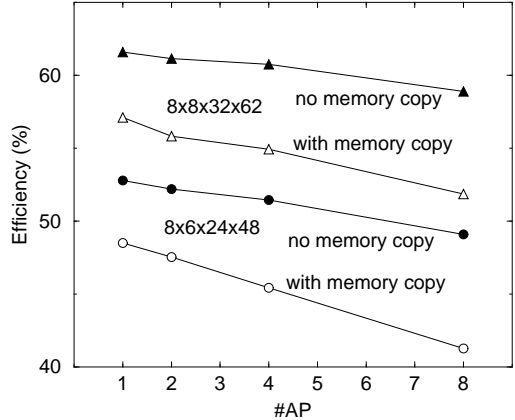}}
\vspace{-12mm}
\caption{MULT efficiency on 1PN versus \#AP.}
\label{fig:MULT-MT}
\vspace{-6mm}
\end{figure}

The cost of automatic parallelization is another important point,
because SMP of vector processors is a distinctive feature of PN. 
Figure~\ref{fig:MULT-MT} shows the efficiency against \#AP for 
$8\times 8\times 32\times 62$ and $8\times 6\times 24\times 48$ lattices.
The micro-tasking parallelization costs 3 to 4\% which is not so high,  
while memory copy to implement boundary conditions 
is relatively heavy, being 4\% for 1 AP and 7\% for 8 AP's.

\subsection{MPI Communications}
In our code one PN issues an MPI\_irsend for a gathered data and an 
adjacent PN issues an MPI\_irecv and then scatters the received data.
This enables us to construct long messages. 
The message size ranges from 0.34MB to 1.64MB, and the 
throughput ranges from 1.62GB/sec to 5.56GB/sec.
These numbers are consistent with a MPI performance report
from the ES Center.

Communication performance drops by 20\% due to buffer copy
for gather/scatter, {\it e.g., } the throughput for the longest
message drops to 4.35GB/sec. 

\subsection{Breakdown of Overheads}
\begin{figure}[t]
\centerline{\epsfxsize=6.7cm \epsfbox{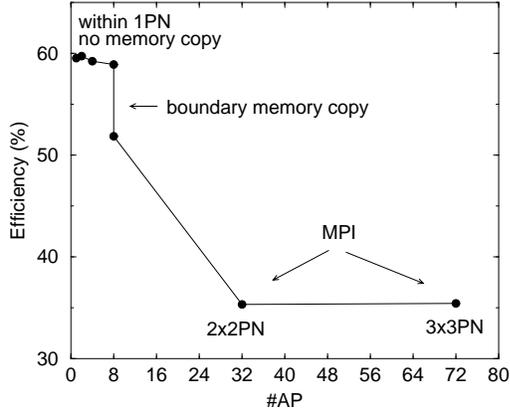}}
\vspace{-12mm}
\caption{Efficiency of MULT versus \#AP.}
\label{fig:MULT-upto72}
\vspace{-6mm}
\end{figure}

\begin{figure}[t]
\centerline{\epsfxsize=6.7cm \epsfbox{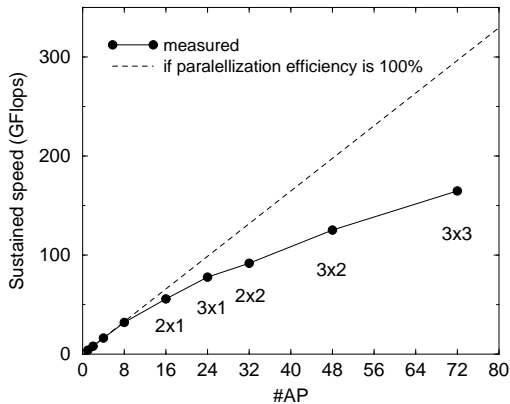}}
\vspace{-12mm}
\caption{MULT sustained speed for a $24^3\times 48$ lattice
versus \#AP.}
\label{fig:scaling}
\vspace{-6mm}
\end{figure}

In order to show how various overheads affect the overall efficiency, 
we show in Fig.~\ref{fig:MULT-upto72} the MULT performance 
%(calculated theoretically)
starting from 1 AP up to 72 AP's (9 PN's). 
The lattice volume per AP is fixed to $4\times 2\times 32\times 62$,  
which corresponds to a $32\times 32\times 32\times 62$ lattice
on a $4\times 4$ PN array. 
The performance of 61.6\% for 1 AP finally drops to 35.4\% for 72 AP's.
The main cause of the drop is a slow speed of
MPI communications which are not overlaid with arithmetic operations,
and secondly the cost of memory copy
in one PN, which together costs 40\% relative to the total 
execution time. The fraction becomes
higher when volume/node becomes smaller; for a $20^3\times 40$
lattice on a $5\times 2$ PN array, 60\% of execution time is
spent for communications and memory copy.

Figure~\ref{fig:scaling} shows the sustained speed in GFlops 
versus \#AP for a $24^3\times 48$ lattice. In this case, 
lattice size is fixed for all measurements. For 1 AP,
4.12GFlops (52\% efficiency) is achived, and  
164.83GFlops for 72 AP's is about 40 times
that for 1 AP. In other words, the parallelization
efficiency is 40/72 $\approx$ 56\%.

\section{Performance of the PHMC Program}\label{sec:PHMC}
\begin{table}[t]
\caption{Profile (\%), performance (GFlops) per 1AP
and total efficiency (\%) of the PHMC program.}
\label{tab:PHMC}
\begin{tabular}{lrrrr}
\hline
size & $20^3\cdot 40$ & $24^3\cdot 48$ & $32^3\cdot 64$ &
$32^3\cdot 62$ \\
node & $5\cdot 2$ & $3\cdot 4$ & $4\cdot 4$ & $4\cdot 4$ \\ 
\hline
copy & 36.9 & 28.6 & 21.4 & 24.6 \\
MULT & 24.8 & 29.4 & 34.8 & 34.1 \\
MLCI & 13.4 & 18.3 & 20.1 & 17.8 \\
BiCG & 9.8 & 9.2 & 8.8 & 8.9 \\
\hline
Perf & 2.01 & 2.29 & 2.87 & 3.18 \\
\hline
Eff  & 24.9 & 28.6 & 35.9 & 39.8 \\
\hline
\end{tabular}
\vspace{-6mm}
\end{table}

Table~\ref{tab:PHMC} shows the profile of the entire PHMC program 
as provided by the ES profiler. 
The arithmetic calculations and 
boundary copy/communications in MULT are the two heaviest routines.
The multiplication of the inverse clover term (MLCI)
and BiCGStab (BiCG) are the third and fourth heaviest, which are 
relatively light. 
Therefore, for the next round of 
simulations on a $32^3\times N_t$ lattice,
we plan to overlay arithmetic operations and communications in
the MULT routine. 

As Table~\ref{tab:PHMC} shows the PHMC program runs on the ES with 
an efficiency of 25--40\% for our four target lattice sizes.
Currently we are executing on a $20^3\times 40$ lattice 
at $m_\pi/m_\rho \approx 0.6$.
The efficiency of 31\% on the ES is comparable 
to that on other machines, 35\% on SR8000/F1 at KEK with 32 nodes,
44\% on VPP5000 at Tsukuba with 8 nodes, and 20\% on CP-PACS
with 500 nodes.

\vspace{2mm}
We would like to thank the ES Center for the approval of our 
project, K.~Itakura for continuous technical support, and
RCNP, Osaka university for some CPU time of SX5 at 
early stage of this project. This work is supported in part by
Grants-in-Aid of the Ministry of Education 
No.15204015.  % Kiban-A

\end{document}